# Utilization of the Transverse Deflecting RF Cavities in the Designed QBA Lattice of the 3 GeV Taiwan Photon Source

Hossein Ghasem[a1], Gwo Huei Luo[b] and Ahmad Mohammadzadeh[c]

[a] *School of Particles and Accelerators, Institute for Studies in Theoretical Physics and Mathematics (IPM), P. O. Box 19395-5531, Tehran, Iran.*

[b] *National Synchrotron Radiation Research Center (NSRRC), Hsinchu 30076, Taiwan.*

[c] *Nuclear Science and Technology Research Institute (NSTRI), P.O. Box 14395-836, Tehran, Iran.*
 *E-mail*: ghasem@ipm.ir

ABSTRACT: A pair of superconducting transverse deflecting RF cavities has been studied in the QBA low emittance lattices of the 3 GeV TPS for generating ultra short X-ray pulses. Three configurations with different locations for the two cavities in a super-period of the TPS ring are investigated. During numerous turns of the electron tracking, the nonlinear effects between the cavities, the energy spread, the momentum compaction factor, and the synchrotron radiation effects are taken into account. The configuration with positioning the RF deflectors between the QBA cells in each super-period as an optimum arrangement gives rise to better quality electron bunches and radiated photon pulses. The FWHM of the radiated photon pulses of about 540 fs with an acceptable intensity is attained by optimizing the compression optical elements of the TPS photon beam line. Furthermore, the effects of the electron bunch length are studied by alternatively employing an accelerating RF cavity operating with 1.1 MV and 3.0 MV, respectively. The operation of the accelerating RF cavity at 3.0 MV improved the intensity of the photon pulses up to 30% and reduced the equilibrium vertical emittance down to 70 pm-rad. The error tolerance for the deflecting cavities, QBA lattice and injection process are also evaluated.

---

[1] Corresponding author.

# Contents

## 1. Introduction

The typical rms bunch length in a storage ring is usually about several tens of picoseconds (ps). This range of pulse duration is useful in numerous experiments but shorter pulses of sub-picoseconds radiation, such as those associated with ultra fast phenomena, would extend the scientific frontier. Many efforts have been made to produce shorter X-ray pulses in synchrotron radiation facilities [1]-[8]. As a third generation light source, Taiwan Photon Source (TPS) would produce about 19 ps-long (rms) X-ray pulses by operating the accelerating radio frequency (RF) cavity at 1.1 MV. The TPS proposed design is made of six super-periods, each consisting of two quadruple-bend achromat (QBA) cells [9]-[10]. The QBA cell consists of two double-bend achromat (DBA) cells of unequal bending lengths associated with the outer and inner dipoles [11]. This type of lattice has some advantages over the double-bend achromat or the double-bend nonachromat by providing a small natural beam emittance and some zero dispersive straight sections. Exploiting these advantages helped us employ the transverse RF deflectors [12]-[18] in our simulation to produce sub-picosecond photon pulses.

Similar to the work performed at Advanced Photon Source (APS) [15] by simulating the deflecting RF cavities in their DBA cells, we have studied the detailed simulation and analysis of X-ray compression using the transverse deflecting RF cavities in our designed QBA lattices of the TPS. Working with the QBA lattices at TPS created more alternatives for locations of the deflectors as compared to the DBA lattices used at APS. Due to the nature of our designed QBA lattices [9], the values of the optical functions and the beam parameters vary for the straight sections, whereas in a DBA lattice design they stay invariant. Therefore, it is essential to investigate different arrangements for positions of the deflecting cavities within the QBA and study the optimum configuration.

This article begins with a brief review of the concept of deflecting cavities in a synchrotron light source. We propose three possible configurations for locations of the cavities in a super-period of TPS and have studied them to find the minimum achievable photon pulses duration. Furthermore, the effects of the non-zero momentum compaction factor and the energy spread on tilted electron bunch as the emittance degradation sources in an ideal machine are investigated at TPS.

By switching on/off the interior sextupoles (the sextupoles that are between the deflectors), the nonlinearities, couplings and chromaticity for all configurations are examined. The synchrotron radiation effects, namely the radiation damping and quantum excitation are taken into consideration during tracking of the electrons to find the equilibrium emittance. Ultimately, the best deflecting structure configuration for generating sub-picoseconds pulse duration is chosen and the radiated pulses are passed through the proposed compression optical elements of the TPS photon beam line. Furthermore, the comparative studies of the electron bunch length for different accelerating voltages for TPS are also carried out and tolerances of the errors associated with simulation of deflecting cavities in the optimum configuration are evaluated.

## 2. Compression system in TPS

### 2.1. Concept of deflecting structures

In a storage ring, a transverse deflecting RF cavity can be employed to produce a correlation between vertical momentum and longitudinal position of the electrons in a bunch. The sinusoidal vertical kick from a deflector leads to a head-tail oscillation of the stored electron bunch in the opposite direction [13]-[14]. In order to confine this coupling in a section of the storage ring, a second deflecting structure must be placed at an integer number of half betatron wavelengths downstream from the first structure. It provides the perfect compensation for all distortions to the longitudinal and transverse motion of the electrons made in the first cavity [12]-[15].

If an insertion device (ID) like an undulator or a wiggler, as a radiator, is placed between the deflecting cavities, the radiated photons would have some correlations among their time, vertical position and vertical slope. In order to reduce the beam size in the radiator source and maximize the angular variation of the slope, it is advantageous to place the ID at locations that are approximately $m\pi$ (m is an integer number) distant from the first cavity in the vertical phase advance. To acquire a shortened X-ray pulse the radiated X-ray is cut by a slit and to enhance this effect an asymmetrically cut crystal [19] can be used. Finally, the standard deviation of the shortened X-ray pulses with Gaussian distribution is given by

$$\sigma_{X\text{-ray}} = \frac{E}{eV\omega_c}\sqrt{\frac{\beta_{yr}}{\beta_{yc}}(\sigma_{y'e}^2 + \sigma_{r'}^2)} \qquad (1)$$

where $\beta_{yr}$ and $\beta_{yc}$ are the vertical beta functions at the radiator and at the cavities, $\sigma_{y'e}$ and $\sigma_{r'}$ are the angular spread due to the uncorrelated vertical beam emittance and the radiation, E is the nominal energy of the electrons, V is the peak deflecting voltage, $\omega_c = h\omega_{RF}$ is the angular frequency of deflecting cavities, h is the harmonic number, and $\omega_{RF}$ is the main angular radio frequency. As seen in Eq. (1), in order to minimize the duration of radiated X-ray pulses, it is beneficial to place the cavities and the ID at locations where the vertical beta function is high and low, respectively. Likewise, increasing the deflecting voltage and the harmonic of the deflectors or both would provide a stronger vertical kick to the electrons (as presented in Figure 1) and significantly compresses the duration of the photon pulses. Moreover, the length of the ID, the radiation wavelength and the divergence of the untilted electron bunch would not have a drastic effect on the pulse duration.

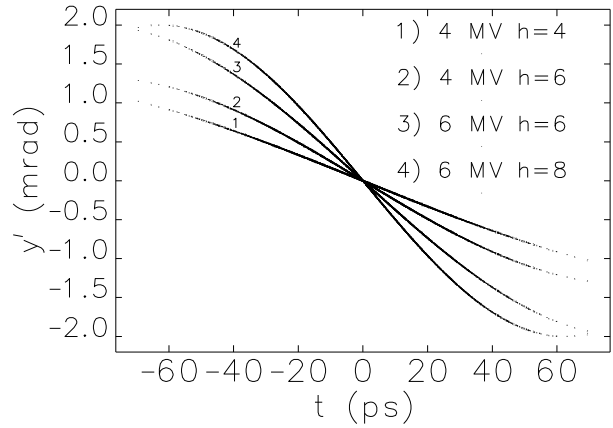

**Figure 1.** The vertical slope of the electrons for various deflecting parameters after passing through the deflecting structure. 10000 electrons per bunch have been employed for tracking using ELEGANT [22] program code. The electron tracking for various h and V illustrates the impact of increasing deflecting parameters.

As seen in Figure 2, for the case where the cavities are placed in the middle of the two QBA cells, the vertical beta function at the cavities and at the ID between the cavities are 1.45 m and 1.37 m, respectively. Furthermore, the divergence of the untilted electron bunch at the ID is 4.67 μrad and the angular spread due to the emitted X-ray having a Gaussian distribution is computed by

$$\sigma_{r'} = \sqrt{\frac{\lambda_r}{2L}} \qquad (2)$$

where $\lambda_r$ is the radiation wavelength and L is the length of the ID. Using the ID parameters given in Table 1, the value of $\sigma_{r'}$ becomes 13.82 μrad. Thus, for E = 3.0 GeV and $f_{RF} \cong 0.499$ GHz, Eq (1). predicts that operating the superconducting deflecting structures in the 8$^{th}$ harmonic at 6.0 MV generates photons with FWHM of around 0.66 ps.

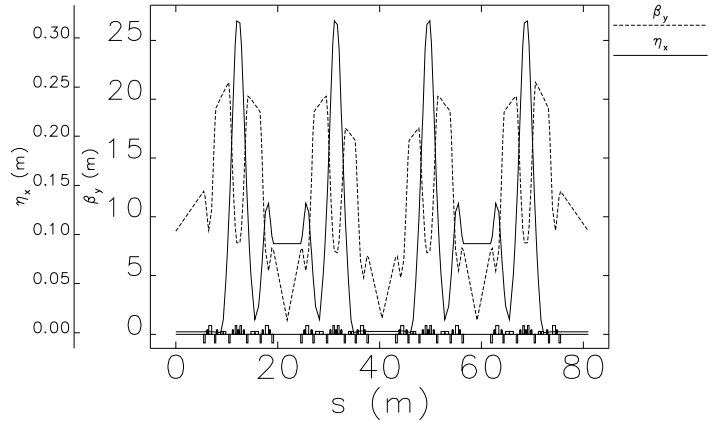

**Figure 2.** The optical functions in a super-period of TPS.

**Table 1.** The ID main parameters in TPS.

| Parameter | Symbol | Value |
|---|---|---|
| Photon energy[KeV] | $E_{ph}$ | 0.811 |
| Current[mA] | I | 400 |
| Number of ID periods | N | 143 |
| Length of ID period[mm] | $L_u$ | 28 |
| Length of ID[m] | L | 4 |
| Gap[mm] | g | 7 |
| $K_{max}$ | - | 2.35 |
| Magnetic field[T] | B | 0.9 |
| Type | - | Hybrid |

## 2.2. Configurations of deflecting structures in TPS

The designed TPS ring is supposed to have six long straight (LS) and 18 short straight (SS) sections each 10.91 m and 5.31 m, respectively [9]. About one third of the ring is composed of straight lines accommodating special devices such as deflecting structures. Ten families of quadrupoles and eight families of sextupoles with mirror symmetry are employed in a super-period. The main TPS parameters are given in Table 2. Fortunately, the vertical phase advance in each QBA lattice is about $2\pi$ which is in favour of the deflecting structures. Three alternative locations for placing the deflectors are designated in a super-period of TPS. To elucidate the effects of deflectors in each alternative location, five watch points are established in the system. Each watch point has the capability of monitoring the property/motion of the beam. As displayed schematically in Figure 3, the watch points are marked (W1) before the first cavity, (W2) after the first cavity, (W3) at the center of the undulator as the radiator, (W4) before the second cavity, and (W5) after the second cavity. All watch points are set on the coordinate mode [22] and in addition, a bunch of 10000 electrons with a Guassian distribution is tracked through each alternative configuration.

In order to produce a rapid angular variation regarding the quantum life time limitation [20]-[21], the cavities are set to be in the 8[th] harmonic of the main RF for all configurations. The description of the three alternative configurations is as follows:

**Table 2.** Main parameters of TPS.

| Parameter | Symbol | Value |
|---|---|---|
| Energy[GeV] | E | 3 |
| Circumference[m] | C | 486 |
| Nat. emittance[nm-rad] | $\varepsilon$ | 3 |
| Tune | $Q_x/Q_y/Q_s$ | $26.27/12.25/3.058\times 10^{-3}$ |
| Nat. chromaticity | $\xi_x/\xi_y$ | -64/-30 |
| Momentum compaction | $\alpha_c$ | $2.712\times 10^{-4}$ |
| Damping times[ms] | $\tau_x/\tau_y/\tau_s$ | 12.94/12.96/6.48 |
| Energy spread | $\delta$ | $8.319\times 10^{-4}$ |
| Energy loss per turn[MeV] | $U_0$ | 0.75 |
| RF gap voltage[MV] | $V_{RF}$ | 1.1 |
| RF frequency[GHz] | $f_{RF}$ | 0.4996540967 |
| Harmonic number | h | 810 |
| Revolution frequency[MHz] | $f_{rev}$ | 0.61728 |
| Bunch length[mm] | $\sigma_l$ | 5.699 |
| Dipole field[T] | B | 1.0479 |
| Dipole length[m] | L | 1 and 1.5 |

First, the deflecting structures are situated in dispersion-free regions at the beginning and at the end of a QBA cell, as shown in Figure 3(a).

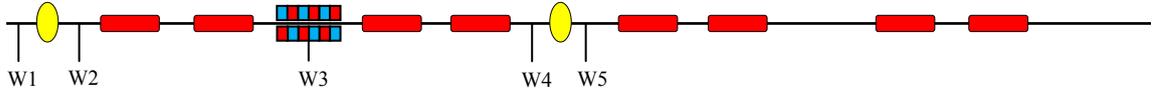

**Figure 3(a)**

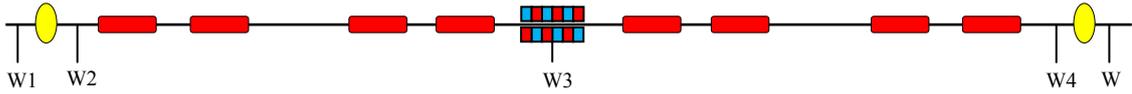

**Figure 3(b)**

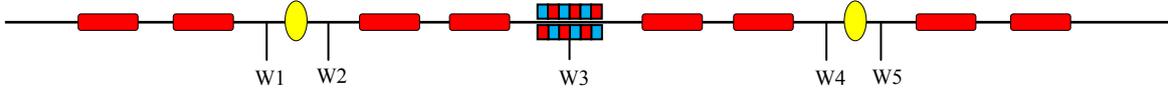

**Figure 3(c)**

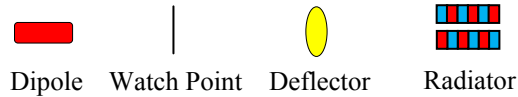

Dipole    Watch Point    Deflector    Radiator

**Figure 3.** The deflecting cavities (a) at the beginning and at the end of a QBA cell (first configuration), (b), at the beginning and at the end of a super-period (second configuration), (c) in the middle of the two QBA cells in a super-period (third configuration). The many quadrupoles and sextupoles that exist in the lattice are omitted from this drawing for clarity.

The systematic fine adjustment of the deflector positions yielded a difference in the horizontal and vertical phase advances of almost 13.71 and 6.28 ($\approx 2\pi$), respectively. The distance from the deflector center to the narrowest aperture free of any elements is D = 5.3 m. For the vertical chamber size of 2A = 9 mm, the upper limit of the deflecting voltage (eV ≤ EA/D) is 2.6 MV.

The nominal vertical positions of the electrons are about several microns when the deflectors are off. However, for a 2.5 MV vertical kick, they are increased by almost a thousand times after a single turn electron tracking employing ELEGANT [22]. This indicates imperfect cancellation of the first kick in this configuration. One can generally expect that the cancellation process would be improved by adjusting the lower vertical kick voltage, but operating deflectors even in low voltages revealed that the kick cancellation process is not as sufficient as required and greatly degraded the transverse emittance of the electron beam. This is mainly related to the QBA lattice functions being different at the position of the two deflectors. The vertical beta function at the first and second cavity is 8.8 m and 1.3 m, respectively. The difference in the vertical beta function at the cavities causes the second kick to be different from the first [12]. Moreover, the electron beam vertical divergence at the second cavity (4.67 μrad) as compared to the first cavity (1.8 μrad) is approximately 2.6 times larger. Therefore, the bunch rotation is not perfectly reversed and the transverse emittance blows up. Since the vertical beta function, the vertical beam size and the vertical beam divergence at the position of the two deflectors are different, the first vertical kick is not perfectly reversed even by reducing the first voltage 2.6 times of the second voltage. Thus, for further improvement of the cancellation process, we reduced the first kick down to 0.5 MV and kept the second kick fixed on 2.5 MV ($V_1 = V_2/5 = 0.5$ MV) and tracked the electron bunch for many turns. As a result, the vertical emittance blow-up became large as compared to the other two configurations (see Figure 7) and the minimum duration of pulses could not be obtained. Thus, we discarded the first configuration from our option list.

In the second configuration, the deflectors are located in the beginning and at the end of the super-period, as shown in Figure 3(b). The difference in horizontal and vertical phase advances of the deflectors are adjusted to be around 27.32 and 12.57 ($\approx 4\pi$), respectively. The vertical beta function at the deflecting cavities and the radiator are 8.93 m and 1.37 m, respectively. Moreover, the dispersion function at the cavities and at the center of the radiator is zero. Although the deflectors have the same lattice functions, the numerous nonlinear elements in the large distance separating them, 78.67 m, greatly affect the tilted bunch and generate error in the vertical divergence of the electrons which in turn degrade the transverse emittance. Furthermore, two out of six long straight sections of TPS would be occupied. In this configuration, the distance from the deflector center to the narrowest aperture free of any elements is 4.28 m and the deflecting voltage should not exceed 3.15 MV. Hence, for producing minimum duration of X-ray pulses, the deflecting voltage is set at 3.0 MV. Finally, the electron tracking after a single turn revealed that the cancellation of the first kick is rather acceptable.

In the third configuration, the deflectors are located in the middle of the two QBA cells in the super-period where two dispersive short straight sections are devoted to the deflectors and the ID is placed at a dispersion free straight section between the deflectors as shown in Figure 3(c). The vertical magnetic fields generated by the deflectors are very weak and thus generate small vertical dispersions which cause small impacts on the vertical emitance. Since, the direction of the deflecting kick is vertical and primarily irrelevant to the horizontal dispersion, the effect of dispersion function can be ignored at the deflectors. The positions of the deflectors are adjusted to yield a difference in horizontal and vertical phase advances of 13.46 and 6.28 ($\approx 2\pi$), respectively. The distance between the deflectors is 36.2 m and the vertical beta function at the deflectors and the ID are 1.45 m and 1.37 m, respectively. Since the distance from the first deflector to the narrowest aperture with no elements is 2.1 m, then the upper limit of the kick voltage is calculated to be around 6.42 MV. To produce a large vertical kick, the deflectors are set at 6.0 MV before the electron tracking. The tracking results indicated that the kick cancellation is as sufficient as the second configuration.

In the 2$^{nd}$ and 3$^{rd}$ configurations, the deflectors produced almost equal vertical correlations at the ID location. Applying Eq. (3) for the 19 picosecond electron bunch length, $\sigma_t$, the electron vertical slope at the radiator for the second and third configurations are calculated to be 1.2 mrad and 0.98 mrad as shown in Figure 4.

$$y' = \sqrt{\frac{\beta_{yr}}{\beta_{yc}}} \frac{eV\omega_c}{E} \sigma_t . \qquad (3)$$

Although the first kick is almost reversed by the second deflector, for both configurations the emittance growth is inevedible. Since the QBA lattice functions at the deflectors are the same, the main sources of emittance degradation and imperfect vertical kick cancellation process are associated with the electrons energy spread and nonlinear elements between the deflectors. The non-zero momentum compaction factor and interior sextupoles affect the amplitude and divergence of the tilted electrons between the deflectors and generate errors in these parameters at the second deflector.

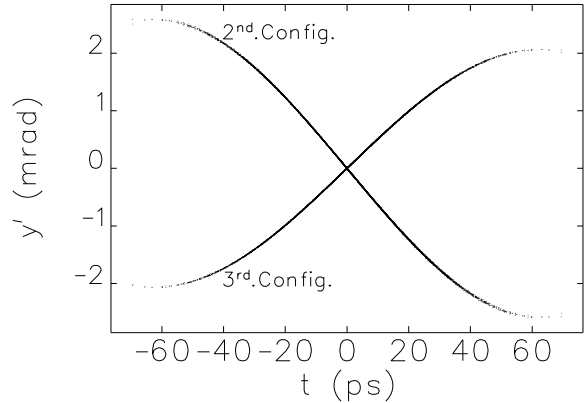

**Figure 4.** The vertical slope of electrons at the center of the ID in the second and third configurations. The deflectors in the 2$^{nd}$ and 3$^{rd}$ configurations are set to 3.0 MV and 6.0 MV, respectively.

## 3. Emittance degradation

In this section, we present detailed analysis and simulation studies of sources of emittance degradation in a perfect TPS machine with the deflecting structures of the 2$^{nd}$ and 3$^{rd}$ configurations. For the extreme case of h = 8 with deflecting voltages of 3.0 MV and 6.0 MV for the 2$^{nd}$ and 3$^{rd}$ configurations respectively, a Gaussian distribution of 10000 electrons per bunch is tracked for a single turn where the synchrotron radiation effects containing radiation damping and quantum excitation are excluded during the tracking. As anticipated, the effects of the emittance degradation sources are diminished when the deflecting voltages were reduced.

### 3.1. Non-zero momentum compaction factor

The non-zero momentum compaction factor effect is present even in an ideal machine where there are no errors and nonlinearities. As presented in Table 2, the energy spread and momentum compaction at TPS ring are fixed at $8.319 \times 10^{-4}$ and $2.712 \times 10^{-4}$, respectively. We can get a sense of why this might matter by computing the differential time-of-flight of the electrons between the deflectors for a fixed energy deviation. The non-zero momentum compaction factor generates a variation of the time-of-flight and thus they have an additional vertical phase term in the second cavity. This can be expressed as follows

$$\psi_{y_2} = n\pi + \psi_{y_1} + \omega_c \Delta t \tag{4}$$

where $\Delta t$ represents the differential time-of-flight of the electrons. The first two terms on the right are the nominal phase advances for an ideal cancellation and the last term is associated with the non-zero momentum compaction factor. For the fraction of the ring between the deflectors, the differential time-of-flight is given by

$$\Delta t = (n/N) \frac{\alpha_c \delta}{f_{rev}} \tag{5}$$

where $f_{rev}$ is the revolution frequency of electrons, $\alpha_c$ is the momentum compaction factor, $\delta$ is the energy spread, and n/N is the fraction of the ring between the deflectors. The fraction n/N is approximately 1/6 and 1/12 for the second and third configurations respectively, and results to $\Delta t_{2nd} \approx 2 \Delta t_{3rd} = 6 \times 10^{-14}$ s. Considering the divergence of the electrons at the second deflector, the error effect in the rms electrons time-of-flight on the emittance blow-up becomes clearer. The electrons differential time-of-flight generates a differential slope error which is evaluated by

$$\sigma_{\Delta y'} = \frac{eV\omega_c}{E} \sigma_{\Delta t}. \tag{6}$$

Since the deflecting voltage for the second configuration is half of the third, the error in rms vertical slope becomes equal for both configurations ($\sigma_{\Delta y'\text{-2nd}} = \sigma_{\Delta y'\text{-3rd}} = 1.5$ μrad). The 1.5 μrad is not negligible while the divergences of the untilted electrons at the deflectors are around 1.8 μrad and 4.9 μrad for the 2nd and 3rd configurations [9] respectively, as given in Table 3. Therefore, using the equation, $\sigma_{y'_2} = \sqrt{\sigma_{y'_1}^2 + \sigma_{\Delta y'}^2}$, the non-zero momentum compaction factor increases the angular spread of $\sigma_{y'_2}$ by a factor of 1.30 and 1.04 for the 2nd and 3rd configurations, respectively. If the nominal values of the beam size are presumed invariant, as given in Table 3, the ratio of the vertical emittance degradation at the second deflector for the configurations could be calculated as $\Delta \varepsilon_{y(2nd)} / \Delta \varepsilon_{y(3rd)} = 6.8$. This large factor makes the 3rd configuration more favourable over the 2nd.

**Table 3.** Approximate values of the vertical beta function, the vertical beam size and the vertical beam divergence at locations of the deflecting structures for the second and third configurations.

| Parameters | Second configuration | Third configuration |
|---|---|---|
| $\beta_y$ (μm) | 8.93 | 1.45 |
| $\sigma_y$ (μm) | 16.4 | 6.6 |
| $\sigma_{y'}$ (μrad) | 1.8 | 4.9 |

### 3.2. Interior sextupoles

Sextupoles are typically required in the storage rings to correct chromatic focusing aberrations and defeat beam instabilities, but they have undesirable effects in the presence of deflectors as well. The horizontal and vertical magnetic fields of a sextupole [20] are as follows

$$B_x = Sxy \quad \text{and} \quad B_y = \frac{S}{2}(x^2 - y^2) \tag{7}$$

where S is the strength of the sextupole. The electrons going through the first cavity receive a large vertical kick such that their vertical amplitudes increase substantially between the deflectors. As shown in Eq. (7), the sextupole magnetic field is nonlinear with respect to the transverse amplitude. Due to this nonlinearity, the large vertical amplitude of the electrons is significantly affected by the interior sextupoles. Since the phase advance and the elimination process of the first vertical kick vary with the transverse amplitude of electrons, nonlinearities of the interior sextupoles lead to an increase in the vertical emittance. Furthermore, because of the interior sextupoles coupling, the horizontal emittance growth occurs as well. The emittance growth due to sextupoles nonlinearities and coupling could be worse than the emittance growth due to the uncorrected chromaticity (partial chromaticity) and in such a case, switching off the interior sextupoles is favoured. Thus, both the on/off operation modes of the interior sextupoles had to be investigated. In the off-mode case, the residual vertical oscillation amplitude of the electrons at the second cavity is given by

$$\langle y_2^2 \rangle^{1/2} = 2\pi \Delta Q \sqrt{\beta_{y_1 c} \beta_{y_2 c}}\, y'_1 \tag{8}$$

where $\beta_{y_1 c}$ and $\beta_{y_2 c}$ are the vertical beta functions at the first and second deflectors, $y'_1$ is the vertical kick of the first cavity, $\Delta Q = -(n/N)\zeta_y \delta$ is the fractional vertical betatron phase error, and $\zeta_y$ is the vertical natural chromaticity of the ring equal to -30 at TPS. The betatron phase errors are calculated as

$$\Delta Q_{2nd} \approx 2\Delta Q_{3rd} = 41.5 \times 10^{-4}.$$

The discrepancy factor of 2 in the betatron phase errors is completely compensated by the difference in their vertical kick voltages of 3.0 MV and 6.0 MV of the deflectors and therefore the dependence of the vertical oscillation amplitude on the vertical beta function becomes the dominant parameter as expressed in Eq. (8). Regarding the beam parameters in Table 3, the residual vertical amplitude for the second configuration calculates to be 6.17 times of the third, ($\sqrt{\langle y_2^2 \rangle_{2nd}} = 6.17 \sqrt{\langle y_2^2 \rangle_{3rd}}$). Therefore, both the non-zero momentum compaction factor and the chromatic effect increase the vertical emittance nominal value of 30 pm-rad to 260 pm-rad and 92 pm-rad for the 2$^{nd}$ and 3$^{rd}$ configurations, respectively. This indicates that the vertical emittance in the 2$^{nd}$ configuration is larger by a factor of 2.8 over the 3$^{rd}$. The ELEGANT simulation results for both the on/off modes of the interior sextupoles are presented in Figure 5.

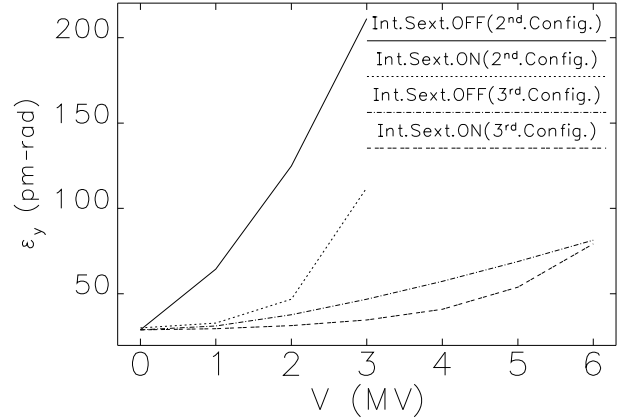

**Figure 5.** The vertical emittance for the 2$^{nd}$ and 3$^{rd}$ configurations after a single pass through the system as a function of deflecting voltage for h = 8. With interior sextupoles off, tracking with momentum spread illustrates the impact of natural chromaticity especially for the second configuration.

In the off-mode case, for the deflecting voltages of 3.0 MV and 6.0 MV, the vertical emittance blow-up almost agrees with the theoretical results obtained above. The observed discrepancy is due to the non-approximated deflecting sinusoidal voltage and the smaller n/N values that were used in the simulation. The linearly approximated deflecting voltage used in the analytical estimation was oversimplified for higher harmonics (h=8). As a result, the vertical emittance blow-up obtained in the simulation is smaller than the theoretical one.

In the on-mode case, as can be seen in Figure 5, for the deflecting voltages of 3.0 MV and 6.0 MV, the vertical emittance increases to 112 pm-rad and 79 pm-rad for the 2$^{nd}$ and 3$^{rd}$ configurations, respectively. As presented in Figure 5, with deflecting voltage of 6.0 MV in the 3$^{rd}$ configuration, the nonlinearity of the interior sextupoles (in the on-mode case) blows up the vertical emittance almost just as much as the partial chromaticity (in the off-mode case). However, with deflecting voltage of 3.0 MV in the 2$^{nd}$ configuration, the nonlinearity of the interior sextupoles blows up the vertical emittance almost half as much as the partial chromaticity. Consequently, working in the on-mode case for both configurations, a lower vertical emittance blow-up is produced especially for lower deflecting voltages. Overall, the 3$^{rd}$ configuration with the interior sextupoles in an on-mode case is more favourable over the 2$^{nd}$ configuration. Furthermore, at higher deflecting voltages of 3.0 MV and 6.0 MV for both configurations, it is anticipated that the on and off-mode vertical emittance blow-up curves would cross each other where the off-mode case would generate a smaller vertical emittance blow-up thereafter.

As far as the horizontal emittance degradation in the third configuration is concerned Figure 6 shows that as the deflecting voltage is increased the horizontal emittance is unaffected in the off-mode case and is blown up to 4.3 nm-rad from 3.0 nm-rad for the on-mode case where the interior sextupoles coupling effect is present.

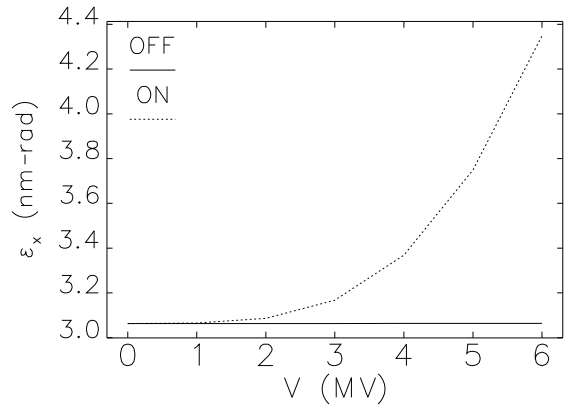

**Figure 6.** The horizontal emittance for third configuration after a single pass through the system as a function of deflecting voltage for h = 8. No increase for the horizontal emittance is seen when the sextupoles are off.

### 3.3. Synchrotron radiation and equilibrium emittance

In addition to the interior sextupoles effects, the synchrotron radiation effects (damping of particle oscillation and excitation of such oscillations) must also be considered during tacking of the electrons. The balance of the two synchrotron radiation effects determines the equilibrium transverse emittance of the electrons in the storage ring. The damping of particle oscillation due to radiation improves the transverse emittance which in turn mitigates the nonlinearity and coupling effects of the interior sextupoles. Since the synchrotron radiation is emitted in quanta of energy, this granular emission effectively provides excitations to the oscillations and thus the effect of quantum excitation degrades the transverse emittance which in turn exacerbates the unwanted interior sextupoles effects. These phenomena can be understood better by tracking the electrons through the system for many turns while considering all the effects during tracking. Since the transverse damping time is almost 13 ms in TPS, the equilibrium transverse emittance can be observed after 8000 turns. Thus, one thousand electrons are tracked for 8500 turns and the ELEGANT simulation results for h = 8 for all three configurations are presented in Figure 7.

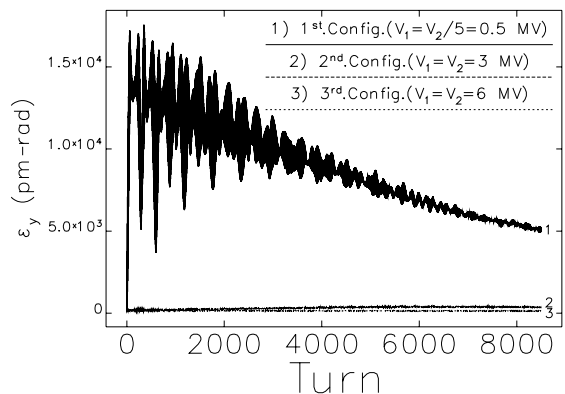

**Figure 7.** The equilibrium vertical emittance for all three configurations. The deflecting voltages of the first and second cavities are equal for both 2$^{nd}$ and 3$^{rd}$ configurations but for the 1$^{st}$ configuration, the first cavity's voltage is one fifth of the second. The interior sextupoles were switched on between the cavities. The deflectors in all configurations were at 8$^{th}$ harmonic of the main RF system.

As predicted above, the 3rd configuration is superior to the 2nd due to the smaller equilibrium vertical emittance. To further compare the two configurations, the electron tracking for many turns at various deflecting voltages is performed. The equilibrium vertical emittance for both the on/off modes of the interior sextupoles is presented in Figure 8.

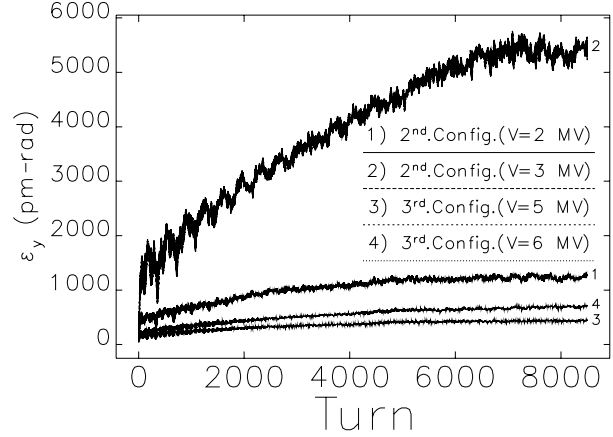

**Figure 8(a)**

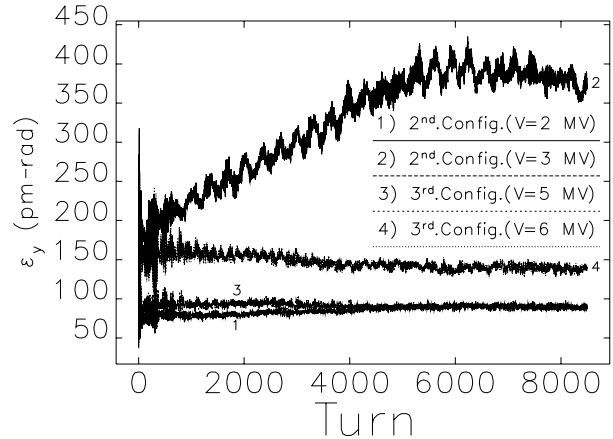

**Figure 8(b)**

**Figure 8.** The equilibrium vertical emittance on successive passes in the 2nd and 3rd configurations of TPS ring for various deflecting voltages and h=8. The interior sextupoles were switched **(a)** off and **(b)** on.

Although the equilibrium horizontal emittance did not change significantly when the interior sextupoles were switched off, but the equilibrium vertical emittance degradation was considerably large, Figure 8(a). For the deflecting voltages of 3.0 MV and 6.0 MV, the eventual vertical emittance in the 2nd and 3rd configurations blows up 12.3 and 4.6 times of its nominal value when the interior sextupoles were switched on as presented in Figure 8(b). Meanwhile, operating the deflecting voltages at 2.0 MV and 5.0 MV, the equilibrium vertical emittance degradation for both configurations becomes comparable. Since increasing the deflecting parameters enlarges the equilibrium vertical emittance and lessens the emitted photon duration, we chose the 3rd configuration and operated the deflecting structures at 6.0 MV and the 8th harmonic of the main RF. Unlike the scenario at APS [15], we decided to keep the interior sextupoles switched on to minimize the equilibrium vertical emittance of the stored electrons and as a result the equilibrium horizontal emittance increased moderately to 3.4 nm-rad. Although, tuning the interior sextupoles [17] for minimization of the transverse emittance was very effective even for high deflecting voltages, but changes to the interior sextupoles were drastic and as a result the dynamic aperture was shrunk extensively and completely unworkable for operation at TPS. Thus we have decided to suspend the interior sexrupoles tuning and have used the interior sextupoles on case in the reminder of our studies.

For comparative purposes, the depicted configuration was run at different voltages and harmonics and the results are presented in Figure 9. It can be observed that the degradation of horizontal emittance from the nominal value is not significant, especially for lower deflecting voltages.

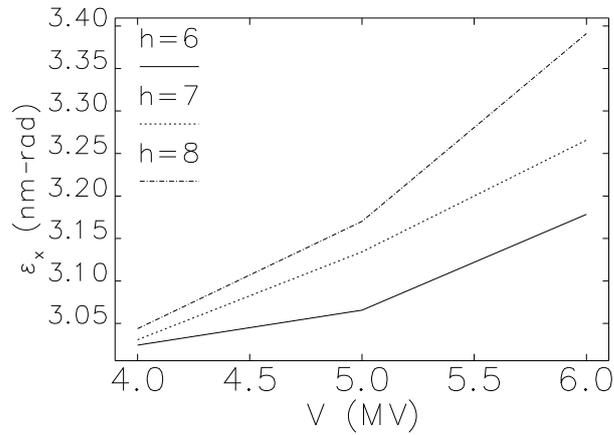

**Figure 9(a)**

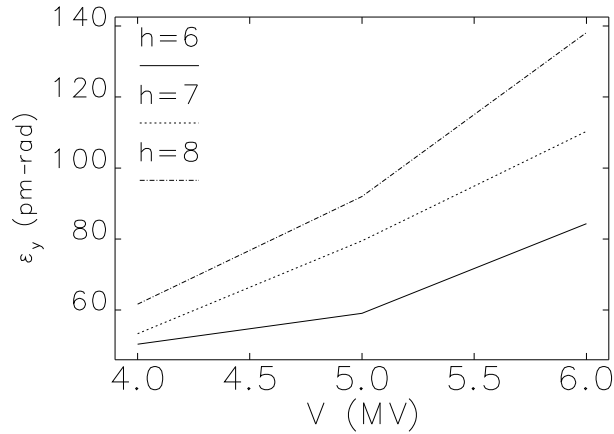

**Figure 9(b)**

**Figure 9.** The eventual (a) horizontal and (b) vertical, emittance in the 3$^{rd}$ configuration versus the deflecting voltages for various harmonics. The interior sextupoles were switched on.

## 4. X-ray compression

Thus far, a Gaussian distribution has been assumed for the shortened X-ray pulses estimation, Eq. (1), while the real distribution of the emitted X-rays has a sinc-function [23], and using

Gaussian approximation results in a somewhat pessimistic analysis of the compression. Thus, a simple method is utilized to simulate the sinc-function distribution for the emitted X-rays. The transverse divergences of the photons emitted from the ID are generated by the following transformations

$$x'_{ph} = x'_{e} + \theta\cos(\varphi) \quad \text{and} \quad y'_{ph} = y'_{e} + \theta\sin(\varphi) \tag{9}$$

where $x'_e$ and $y'_e$ are the transverse divergences of the electrons in the ID location, $\theta$ and $\varphi$ are the random numbers that are sampled with regard to the sinc-function and a uniform distribution, respectively. After generating the sinc-function distribution, the radiated photons are tracked in the TPS photon beam line which is composed of a 60 m long drift space, a slit and an asymmetrically cut crystal. When the photons are drifted a long distance, the pulses can be shortened by slicing the photons using a slit. In addition, an asymmetrically cut crystal based on different angles of incidence and diffraction is employed to induce a variation in time-of-flight of the photons for acquiring a shorter pulse duration in a special plane as shown in Figure 10.

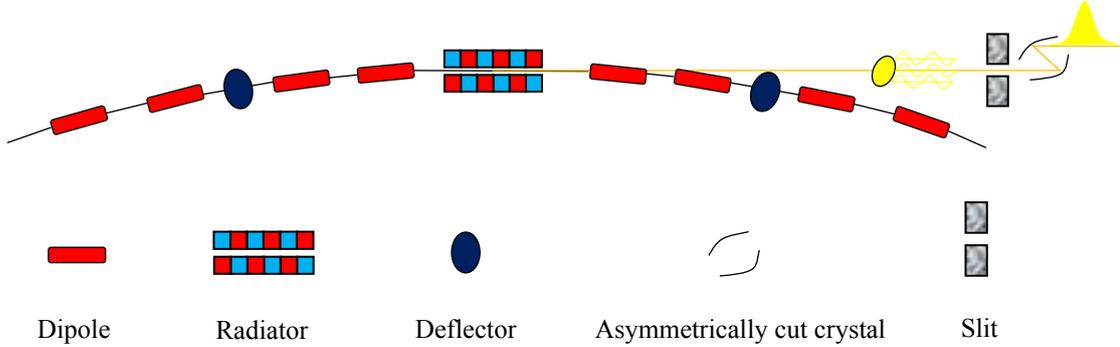

**Figure 10.** The schematic view of compression system in TPS composing of a pair of deflecting structures in middle of two QBA cells in a super-period, a slit and an asymmetrically cut crystal in the photon beam line. The emitted photons from the ID are passed the long drift and sliced by the slit for generation of short X-ray pulses. The duration of radiated pulsed are minimized by optimizing the asymmetrically cut crystal where the lattice planes have an angle with regard to the surface.

For simulation of the asymmetrically cut crystal effect, a simple matrix called "EMATRIX" element of ELEGANT [22] is utilized in which all entries outside the main diagonal are zero except the $R_{53}$ element. Four watch points are established before and after the drift space and the crystal, to monitor the phase space of the photons. Using the watch points, the divergences of radiated photons are monitored step by step while the slit opening is kept at 10 mm, Figure 11.

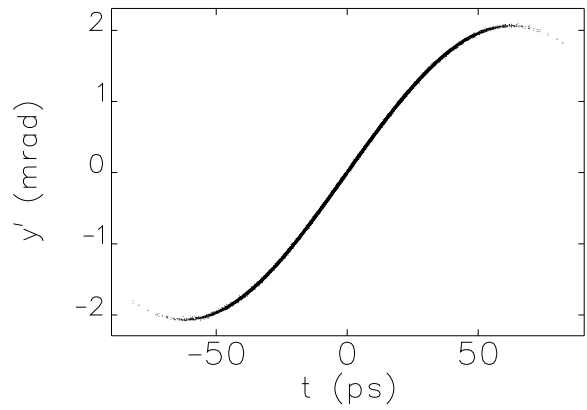

**Figure 11(a)**

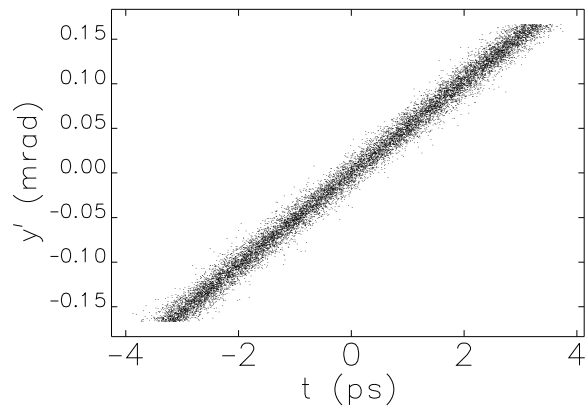

**Figure 11(b)**

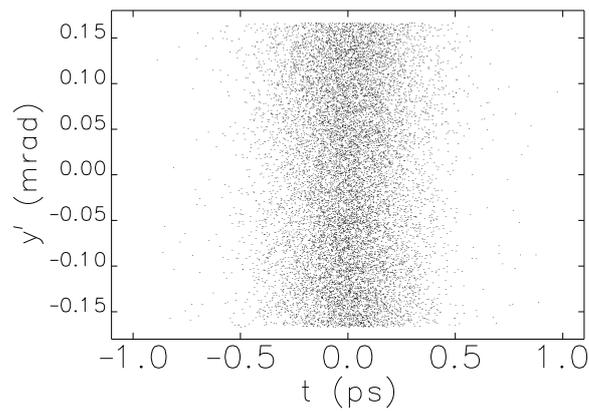

**Figure 11(c)**
**Figure 11.** The vertical divergence of photons in the beam line, **(a)** after the 60 m drift space prior to the slit, **(b)** after the slit and prior to the crystal and **(c)** after the crystal.

It is observed that since the radiation induced divergence is relatively small after the 60 m drift, the vertical divergence of the radiated photons for $8^{th}$ harmonic and 6.0 MV, Figure 11(a), is comparable with the divergence of the electrons at the center of the ID, Figure 4. The radiated photons after drifting the 60 m distance are cut by the slit in the beam line. The minimum duration of the radiated pulses is achieved by finding the optimized value of the $R_{53}$ element according to the equation $\Delta s = R_{53}\Delta y$, where $\Delta s$ and $\Delta y$ are the longitudinal and vertical differential positions of the photons. The photons' vertical slopes before and after the crystal are shown in Figure 11(b) and Figure 11(c), respectively. The FWHM of the radiated photons reflected from the crystal for various slit sizes, different deflecting voltages and harmonics is shown in Figure 12.

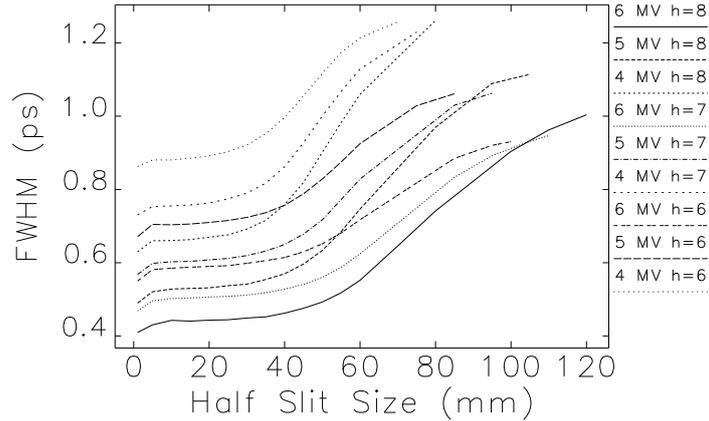

**Figure 12.** The FWHM of central X-ray pulses as a function of the slit-half-height for various deflecting voltages and harmonic numbers of 6, 7 and 8.

This indicates that for the extreme case of h = 8 and V = 6.0 MV, the minimum FWHM of the radiated pulses for a small slit size is around 0.42 ps with low intensity. This value almost agrees with the analytical result of 0.66 ps. The discrepancy is partially due to the Gaussian distribution assumption used for the photons in the theoretical calculations. Additionally, in the analytical evaluation the angular variation was assumed linear, while for the simulation, the element RFTM110 [22] is used where it generates a sinusoidal angular variation. This is an oversimplified assumption especially for higher harmonics.

As far as the intensity of the reflected photons is concerned, the photon transmission versus the slit size for various deflecting parameters is obtained and as presented in Figure 13 for around a 70% photon transmission, the minimum pulse duration (for h = 8 and V = 6.0 MV) is around 0.54 ps. This is associated with the electron bunch horizontal and vertical emittances of 3.4 nm-rad and 138 pm-rad, respectively. Likewise, for V = 4.0 MV, associating with the electron bunch horizontal and vertical emittances of 3.05 nm-rad and 61 pm-rad, a pulse duration of around 0.6 ps is generated. As a result of sacrificing this 20% difference in pulse duration, the vertical emittance is improved more than 50%.

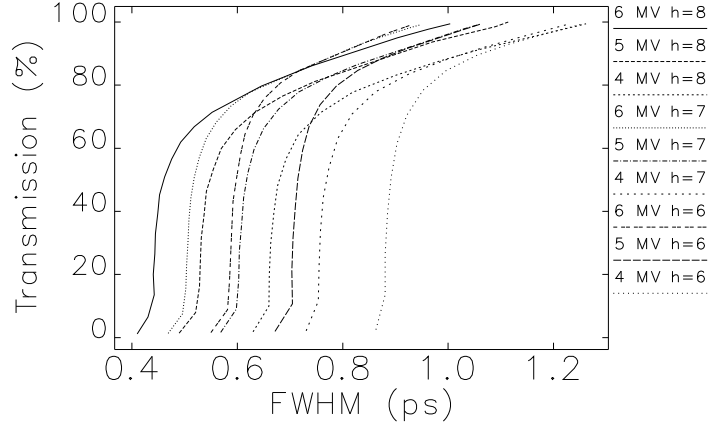

**Figure 13.** The transmission of photons through the silt versus duration of the radiation, for various deflecting structure parameters.

Therefore, the optimum parameters of deflecting structures can be chosen regarding the vertical emittance degradation of electrons (Figure 9), the duration of radiated photons (Figure 12), and the photon transmission through the slit (Figure 13).

## 5. Bunch length effects

In this section we study the effects of any changes of the longitudinal rms bunch length on the electron beam vertical emittance and the radiated photons. The longitudinal rms bunch length in TPS for 1.1 MV accelerating normal RF cavity would be 19 ps which can be manipulated by the RF voltage whereby increasing the voltage decreases the length. The bunch length is reduced down to 10 ps when the accelerating RF cavity is run at the voltage of 3.0 MV instead of 1.1 MV. The 3.0 MV is used in this simulation to reduce the rms bunch length considerably. The extreme case of V = 6.0 MV and h = 8 for the deflecting structures is assumed.

### 5.1. Equilibrium transverse emittance

The electrons in a segment at a longitudinal distance, z, from the bunch center after passing the first deflector receive a sinusoidal vertical kick [13]-[14] which is given by

$$y' = \frac{eV}{E}\sin(\omega_c t) \approx \frac{eV\omega_c z}{Ec}. \qquad (10)$$

Thus, for both 19 ps and 10 ps electron bunches the vertical slopes of 0.95 mrad and 0.50 mrad are obtained respectively as shown in Figure 14. As presented, the sinusoidal formation of vertical kick is clear for the lower gap voltage and thus the vertical amplitude of the electrons for the higher RF voltage becomes smaller than for the lower RF voltage. Consequently, the nonlinear and coupling terms of the sextupole magnetic fields do not affect the kicked electrons of shorter bunch as much and under such circumstances keeping the interior sextupoles in an on-mode would be beneficial.

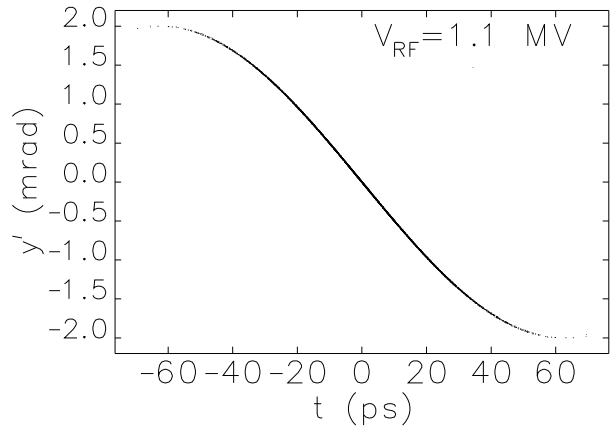

**Figure 14(a)**

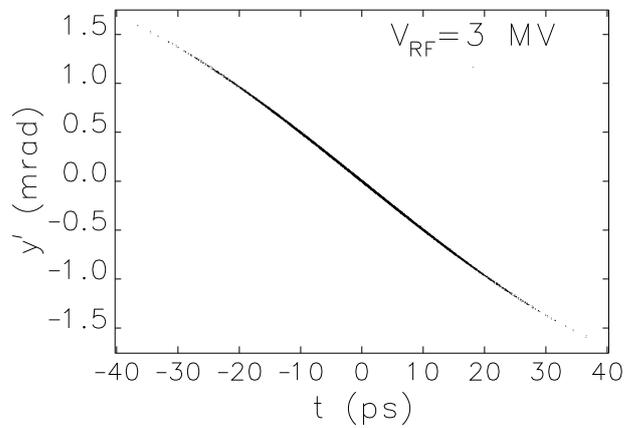

**Figure 14(b)**
**Figure 14.** The vertical slope of electrons for **(a)** 1.1 MV and **(b)** 3.0 MV, accelerating RF cavity operation in TPS ring. The deflectors were set at h = 8 and V = 6.0 MV.

The equilibrium transverse emittance as a function of turn for both operating voltages of 1.1 MV and 3.0 MV is shown in Figure 15.

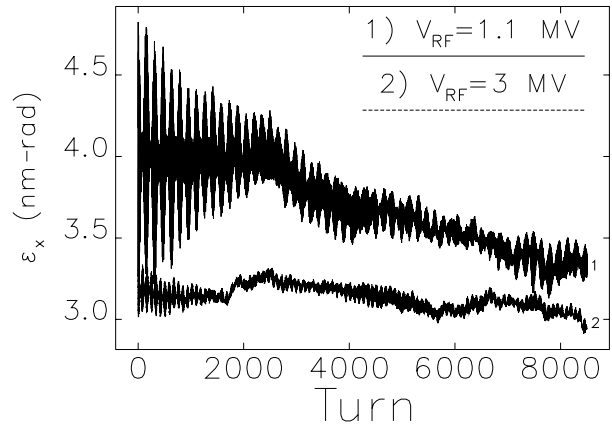

**Figure 15(a)**

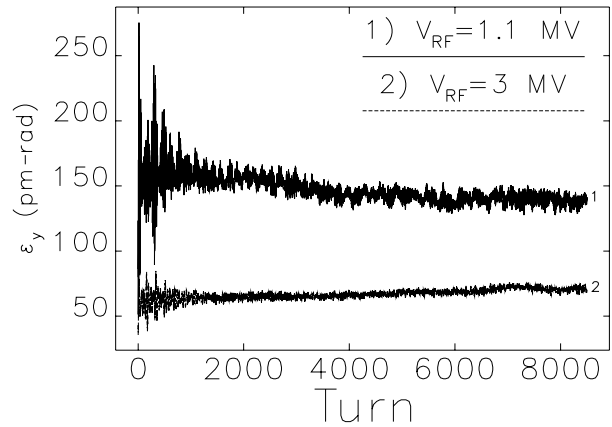

**Figure 15(b)**
**Figure 15.** **(a)** The horizontal and **(b)** vertical, emittance degradation on successive passes for both operations of the accelerating RF cavities in TPS.

As mentioned before, with 1.1 MV accelerating voltage, the 6.0 MV deflecting kick blows up the equilibrium vertical emittance to around 138 pm-rad while with 3.0 MV accelerating voltage it decreases to half as much. The equilibrium transverse emittance drop versus the accelerating voltage is plotted in Figure 16. As anticipated, increasing the RF voltage reduces the bunch length which in turn decreases the emittance. As a result, operating the accelerating RF cavity at higher voltages is beneficial in view of emittance degradation.

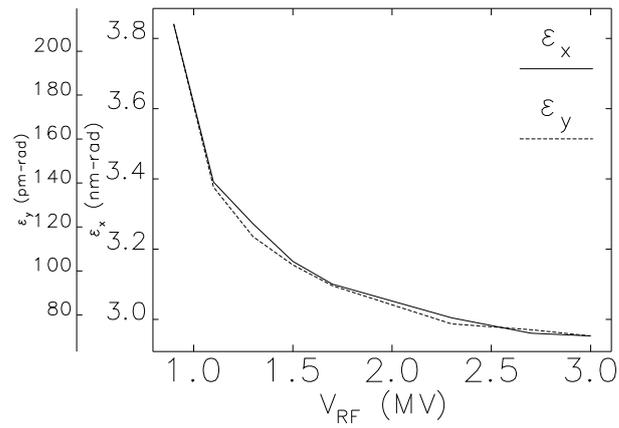

**Figure 16.** Reduction of the eventual transverse emittance by increasing the accelerating RF voltage.

## 5.2. Radiated photons

As mentioned before, the characteristics of the radiated photons from the ID are associated with the electron beam characteristics. And since the density of the electrons at a fixed distance from the bunch center for the 3.0 MV accelerating voltage is more than the 1.1 MV, a higher transmission of radiated photons through the slit is anticipated for the 3.0 MV. The photon transmission versus the slit size and the FWHM of radiated pulses are plotted in Figure 17.

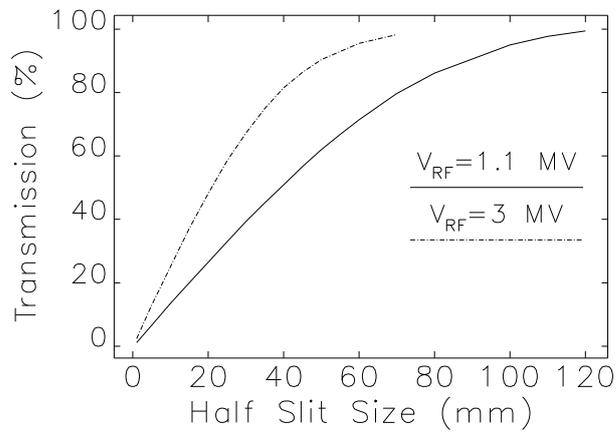

**Figure 17(a)**

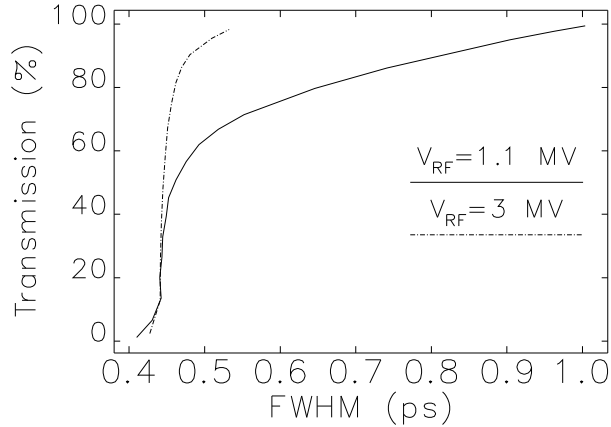

**Figure 17(b)**
**Figure 17.** Transmission of the photons through the slit in the TPS photon beam line versus **(a)** half-slit-size and **(b)** FWHM of the radiated pulses, for various accelerating voltages.

It indicates that for $V_{RF}$ = 3.0 MV and slit size of 140 mm associating to FWHM of 0.54 ps, the photon transmission is approximately 30% more than the $V_{RF}$ = 1.1 MV. Moreover, for a complete photon transmission a difference of around 500 fs between the minimum FWHM of the photon pulses is observed. Therefore, the shorter electron bunch is more favourable for generating ultra short X-ray pulses using deflecting structures.

## 6. Tolerance of errors

Attempts are made to cancel the first kick at the second deflector and reduce the leakage of vertical emittance between the cavities. Typically, even for an ideal storage ring, perfect cancellation does not happen at the second cavity as seen before. The degradation of the equilibrium emittance is mainly related to the nonlinearities and coupling of the interior sextupoles, non-zero momentum compaction factor that generates a variation of the time-of-flight, energy spread, radiation damping and quantum excitation. Since errors are characteristics of a real machine, any errors associated with the compression system for the selected configuration should be considered and their tolerances must be evaluated. The simulation of the main errors due to the deflecting structures, the QBA lattice and injection system are presented and their tolerances are evaluated. The errors related to deflecting voltage, deflecting RF phase and the rolling of cavities are primarily explained. The errors of the QBA lattice functions such as the vertical beta function at locations of the cavities and the vertical phase advance difference between the cavities are taken into consideration. We have also simulated the errors in the centroid of the electron bunch at the presence of compression systems. The offset in the centroid of the electron bunch from the nominal orbit could be created by a mismatch injection of the electrons, referred to as injection error. Any offset can produce imperfect cancellation that leads to degradation of the equilibrium transverse emittance. Once again, the cavities are assumed to operate in the $8^{th}$ harmonic of main RF system throughout the section.

## 6.1. Deflecting voltage

Any error in the adjusted deflecting voltage of the cavities has a direct effect on the vertical kick. Voltage deviation from the nominal value causes the second kick to be different from the first. The rms slope error due to this deviation is given by

$$\Delta y' = \frac{\omega_c \sigma_t}{E} e \Delta V. \tag{11}$$

The error in the vertical divergence of the electrons leads to an imperfect cancellation and degradation of vertical emittance. For simulating this error, the voltage of the first cavity is fixed at 6.0 MV to generate the minimum duration of X-ray pulses and the second deflecting voltage is set around this value. The effect of voltage deviation on the equilibrium vertical emittance is shown in Figure 18.

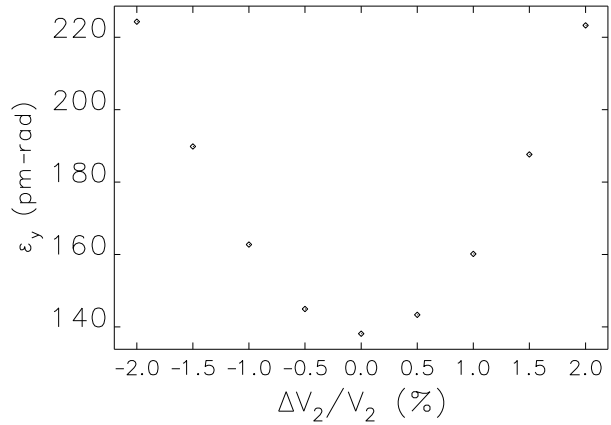

**Figure 18.** The eventual vertical emittance versus the relative voltage error in the second cavity.

It indicates that the impact on the emittance for the relative voltage error of a fraction of a percent is modest. Requiring a voltage error of under 0.5% is prudent. We have also found that the equilibrium horizontal emittance is not very sensitive to this error and it can be neglected.

## 6.2. Deflecting RF phase

The second aspect of deflecting structure errors is related to the RF phase of the cavities. In order to maximize the vertical kick using RFTM110 as an element of ELEGANT, it is essential that the first and second deflectors operate on 90 and 270 degrees RF phases, respectively. A coupled variation of the RF phase from these values only produces a smaller kick and thereby a smaller equilibrium emittance, but minimum pulse duration is not attainable. An uncoupled variation of the RF phase leads to an imperfect vertical kick cancellation. In order to simulate the uncoupled RF phase error, in a manner similar to the voltage error, the first RF phase is fixed at 90 degrees and the second is set around 270 degrees. The tracking results, as shown in Figure 19, demonstrate that the eventual vertical emittance is sensitive to the RF phase indicating that the uncoupled RF phase error should not be far from zero.

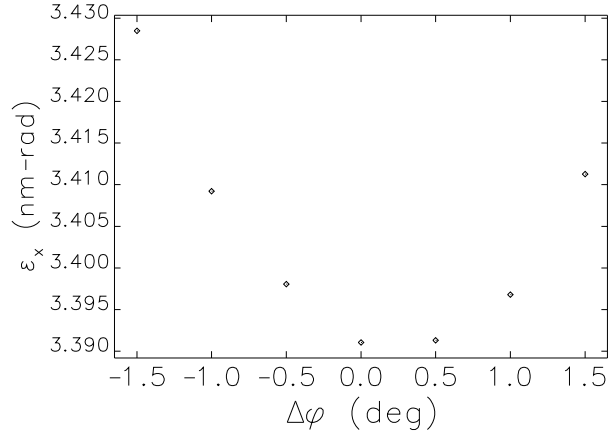

**Figure 19(a)**

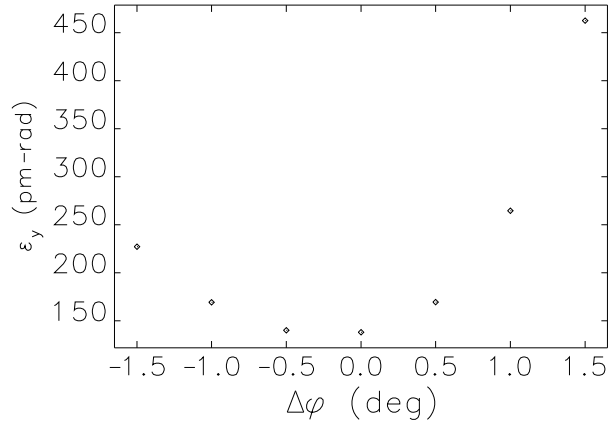

**Figure 19(b)**

**Figure 19.** The eventual **(a)** horizontal and **(b)** vertical, emittance as a function of uncoupled phase error.

Additionally, as anticipated and similar to the voltage error, it is observed that the horizontal emittance is not very sensitive to this error (see Figure 19(a)).

**6.3. Rolling of deflecting structures**

In order to generate a vertical kick, the deflecting structure should be operated in $TM_{110}$ mode [24]. The transverse magnetic fields of this operation mode with the leading order in a simple pillbox cavity are given as follows

$$B_x \approx \frac{E_0 \sigma_x \sigma_y \omega_c^2}{8c^3} \cos(\omega_c t) \quad \text{and} \quad B_y \approx \frac{E_0}{2c} \cos(\omega_c t) \quad (12)$$

where $E_0$ is the electric field, c is the speed of light and $\sigma_x$ and $\sigma_y$ are the horizontal and vertical beam sizes of about 116.6 μrad and 6.58 μrad, respectively. According to Eq. (12), the vertical component of the magnetic force is much smaller than the horizontal one. Therefore, the cavities are rolled 90 degrees using the TILT option [22] of RFTM110 to simulate the desired vertical kick.  In this case, the horizontal force stays negligible and any horizontal emittance blow-up due to the reinforcement of the horizontal force can mainly be associated with rolling of the deflectors.  The errors in the girders or in the installation of deflectors may generate a rotation around the longitudinal axis.  Deflectors undergo coupled and uncoupled types of rolls.  Both deflectors are rotated in the same direction to simulate a coupled roll.  The uncoupled roll is simulated by rolling the second cavity around 90 degrees while the first one is fixed.  Degradation of the transverse emittance as a function of turn for the two types of rolls with different degrees is shown in Figure 20.

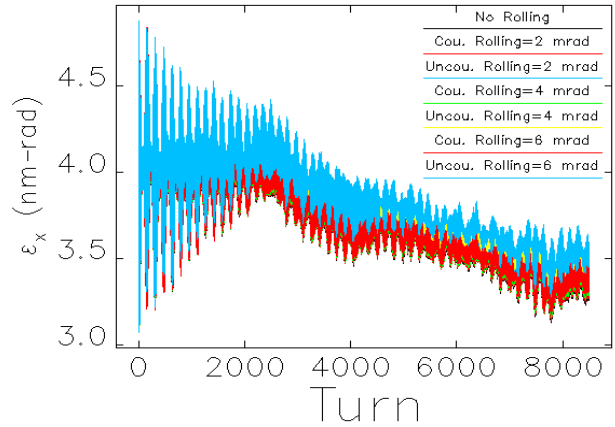

**Figure 20(a)**

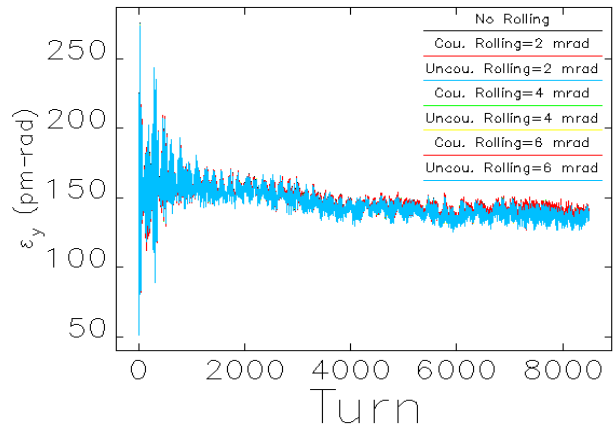

**Figure 20(b)**

**Figure 20.** The **(a)** horizontal and **(b)** vertical, emittance degradation versus the number of turns for both coupled and uncoupled rolling of the deflectors.  The "Cou. Rolling" data is for coupled rolling of deflectors and "Uncou. Rolling" data is for uncoupled rolling mode.

The vertical emittance which overlaps for various rolls (Figure 20(b)) reveals that the degradation of the vertical emittance is insensitive to the rolls.  The horizontal emittance blow-up of up to 6 mrad is not large as seen in Figure 20(a). Therefore, the roll of the deflectors is not as significant as other associated errors and it is easily maintained under a few milliradians using present-day alignment techniques.

### 6.4. Vertical beta function error

The vertical beta function error is exclusively associated with the lattice. As mentioned in Section 2, the divergence of the electrons at integer $\pi$ vertical phase advance downstream from the first cavity is given by Eq. (3). As it can be seen in the equation, any discrepancy between the vertical beta function at the deflectors generates different divergences for the electrons which in turn leads to emittance degradation. This accounts as one of the main reasons for an imperfect cancellation for the first configuration which led to an exclusion of this configuration. The beam line steering, power supply drift and misalignment can be the sources for this discrepancy. The vertical beta function at the locations of the deflectors in the QBA lattice is 1.45 m and a simple simulation method is employed to change this value at the second deflector to assess the error tolerance. Since the vertical betatron phase advance between the deflectors in the third configuration is around $2\pi$, the transfer matrix [25] from the first deflector to the second is given by

$$T = \begin{bmatrix} \sqrt{\dfrac{\beta_{y2}}{\beta_{y1}}} & 0 \\ 0 & \sqrt{\dfrac{\beta_{y1}}{\beta_{y2}}} \end{bmatrix} \quad (13)$$

where $\beta_{y1}$ and $\beta_{y2}$ are the vertical beta functions at the first and second deflectors, respectively. It motivated us to simulate this error by applying a simple diagonal matrix with a determinant of one, EMATRIX [22], prior to the second deflector. The matrix elements $R_{33}$ and $R_{44}$ are set different than one to change the vertical beta function at the second deflector and the inverse of this matrix is employed after the second deflector to undo the perturbation. All other entries in the main diagonal are set to one. As shown in Figure 21 the simulation results indicate that this error must be kept less than 1%.

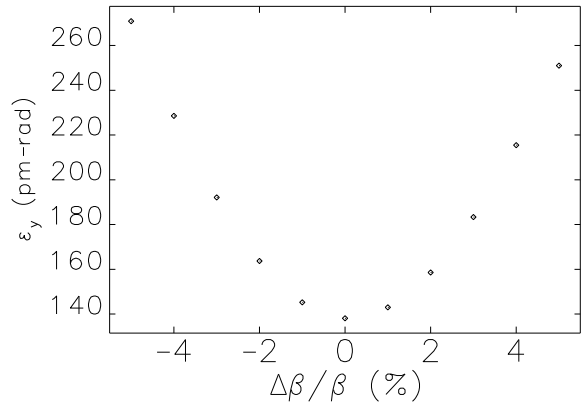

**Figure 21.** The eventual vertical emittance as a function of beta function difference between the two deflectors.

### 6.5. Vertical betatron phase advance error

The second error associated with the lattice arises from not exactly an integer $\pi$ vertical phase advance difference between the deflectors. It causes the vertical position and divergence of the electrons at the second cavity to be different from the first and as a result the first kick is not compensated by the second deflector. As mentioned in Section 2 for the third configuration, the second cavity is at 6.28 vertical phase advance downstream from the first cavity. The dependency of the beta function on the phase advance [25] has made the exact evaluation of this error difficult. Therefore, we moved the deflectors closer together or farther apart symmetrically where the beta functions stayed identical at the two deflectors and only the phase advance changed. Figure 22 shows the equilibrium vertical emittance relative sensitivity to the normalized vertical phase advance error.

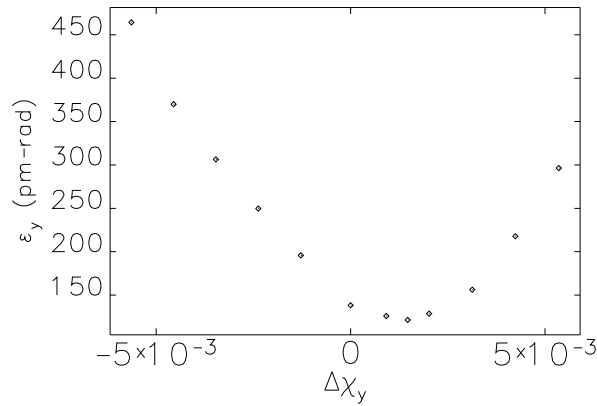

**Figure 22.** The eventual vertical emittamce as a function of $\Delta\chi_y$ defined as a phase advance difference $\Delta\psi_y$ in the equation of $\Delta\chi_y = (\Delta\psi_y/2\pi) - 1$.

The minimum point in the figure is slightly offset from the phase advance of $2\pi$. This presumably results from the use of canonically integrated quarupoles in our simulation which does not give the exact phase advance from the lattice functions. The results indicate that the vertical phase advance error of up to 0.1% can be ignored.

### 6.6 Error of injection system

Typically, a long straight section for injection of the electron bunch is used in storage rings. Any error in the injection kickers, such as excitation amplitude, the uniformity of the metallic coating inside the ceramic chamber or time jitter associated with either construction or installation can cause a mismatch between the bumps of the kickers during the injection and lead to an offset in the centroid of the electron bunch. The centroid of the electron bunch oscillates around the target orbit because of the leakage of the bump of the kickers. The transfer offset in the electron bunch centroid is strongly affected by the deflecting structures and we must model the offset to find the tolerance of this error. When the centroid passes through the ring optical elements in the non-nominal orbit, it receives a kick from the quadrupoles according to the dipole field term. Moreover, the harmful effects of nonlinearity and coupling of the interior sextupoles are better sensed by the centroid. To model this error, equal offsets in both horizontal and vertical directions ($\delta_x = \delta_y$) from the target orbit are considered to simulate the offset as shown schematically in Figure 23.

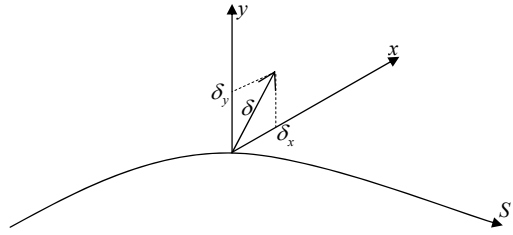

**Figure 23.** The schematic offset of the bunch centroid from the nominal trajectory in the transverse plane.

For the extreme case of 6.0 MV and $8^{th}$ harmonic, the equilibrium transverse emittance dependency on the centroid deviation is shown in Figure 24.

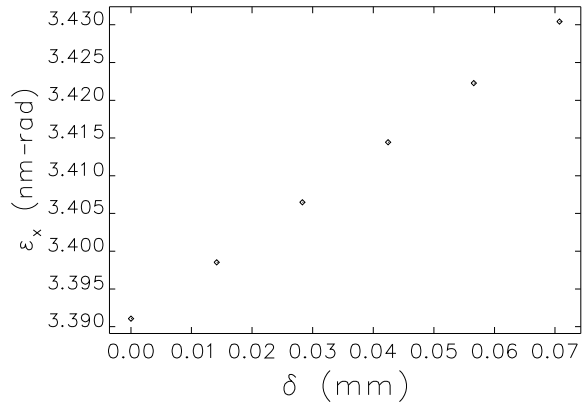

**Figure 24(a)**

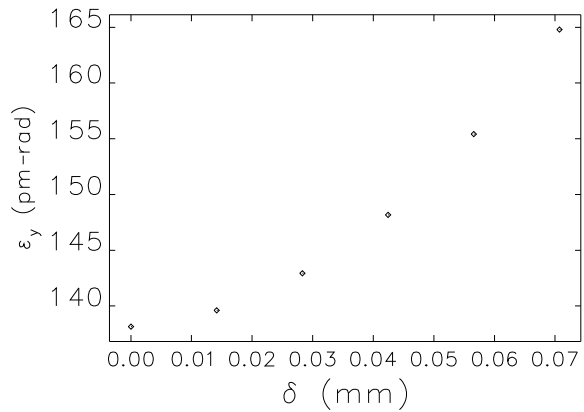

**Figure 24(b)**

**Figure 24.** The eventual **(a)** horizontal and **(b)** vertical, emittance as a function of the radial distance, $\delta = \sqrt{\delta_x^2 + \delta_y^2}$, from the nominal orbit in the transverse plane.

It indicates that after many turn tracking, the eventual equilibrium horizontal emittance degradation is rather insensitive to the offsets. However, the eventual equilibrium vertical emittance increasingly blow up for offsets of more than 20 μm.

## 7. Summary and conclusion

We have studied the transverse deflecting RF cavities in the QBA lattices at TPS to generate ultra short X-ray pulses. Three configurations for locations of the deflectors were investigated. The results showed that the first configuration due to the imperfect vertical kick cancellation of electrons by the second deflector had been excluded and the third configuration was preferred over the second due to the lower eventual equilibrium emittance. It was observed that the equilibrium vertical emittance in the $2^{nd}$ configuration for the $8^{th}$ harmonic at 3.0 MV was 2.6 times of the $3^{rd}$ configuration for the same harmonic at 6.0 MV. The eventual equilibrium transverse emittance for the $3^{rd}$ configuration for various deflecting parameters is presented in Figure 9. In the electron bunch tracking, nonlinearities and coupling of interior sextupoles, momentum compaction factor and synchrotron radiation effects were taken into account to find the eventual equilibrium transverse emittance. Furthermore, the tolerances of the errors associated with the deflectors in the $3^{rd}$ configuration were evaluated.

By optimizing the compression optical elements in the 60 m TPS photon beam line, the lowest achievable pulse duration with low intensity was found to be around 0.42 ps. The FWHM and transmission of the radiated photons for various deflecting parameters are presented in Figure 12 and Figure 13. The FWHM of about 0.54 ps for around a 70% transmission of radiated photons through the slit was obtained.

Consequently, as presented in Figure 9, Figure 12 and Figure 13, the deflecting parameters can be set according to the desired experiment requirements. When the transmission and duration of radiated pulses are in focus it is beneficial to set the deflecting parameters to the highest values regardless of large beam emittance blow-up. However, when the beam emittance is important the deflecting parameters should be set to the lower values.

## Acknowledgments

I would like to thank M. Borland and M. H. Wang for their helpful discussions. I am also grateful to the National Synchrotron Radiation Research Center (NSRRC) for its hospitality and cooperation.